\documentclass[a4paper,11pt]{article}
\pdfoutput=1
\usepackage{pos}
\usepackage{amsmath}
\usepackage{physics}
\usepackage{graphicx}
\usepackage{tikz}
\usetikzlibrary{decorations.markings}
\usetikzlibrary{shapes.misc}
\usepackage[caption=false]{subfig}
\usepackage{hyperref}

\title{Estimating excited states contamination of $B \to \pi$ form factors using heavy meson chiral perturbation theory}
\ShortTitle{Excited states contamination in $B \to \pi$ form factors}

\author[a]{Oliver B\"ar}
\author*[a]{Alexander Broll}
\author[a,b]{Rainer Sommer}

\affiliation[a]{Humboldt-Universit\"at zu Berlin,\\
Newtonstrasse 15, 12489 Berlin, Germany}
\affiliation[b]{Deutsches Elektronen-Synchrotron DESY,\\
Platanenallee 6, 15738 Zeuthen, Germany}

\emailAdd{obaer@physik.hu-berlin.de}
\emailAdd{alexander.broll@physik.hu-berlin.de}
\emailAdd{rainer.sommer@desy.de}

\abstract{Using Heavy Meson Chiral Perturbation Theory (HMChPT), the $B^* \pi$ excited states contamination of the $B \to \pi$ vector form factors is computed to NLO in the chiral expansion and in the static limit. The results suggest that the excited states for $h_\parallel$ are of the order of a few percent whereas $h_\perp$ receives large negative contributions and thus might be significantly underestimated in lattice simulations.}

\FullConference{%
The 39th International Symposium on Lattice Field Theory,\\
8th-13th August, 2022,\\
Rheinische Friedrich-Wilhelms-Universit\"at Bonn, Bonn, Germany
}

\begin{document}
\tikzset{->-/.style={decoration={
  markings,
  mark=at position .5 with {\arrow{>}}},postaction={decorate}}}
  
\tikzset{cross/.style={cross out, draw=black, minimum size=2*(#1-\pgflinewidth), inner sep=0pt, outer sep=0pt},
cross/.default={1pt}}

\maketitle

\section{Introduction}

Chiral Perturbation Theory (ChPT) has proven to be a reliable tool for computing the excited states contamination of lattice correlators. For example,  in \cite{Bar:2018xyi,Bar:2019gfx} the excited states of the induced pseudoscalar form factor of the nucleon $\tilde{G}_p$ were computed in baryon ChPT and were found to be a possible explanation for the discrepancy between lattice simulations and experimental results for $\tilde{G}_p$. It can even help to test the validity of lattice methods to eliminate excited states, see \cite{Bar:2019igf}.

Semileptonic decays of $B$ mesons are an active field of research with the aim of determining CKM matrix elements to a high precision. For $B\to \pi \ell \bar{\nu}$, this requires the computation of the matrix elements $\mel{\pi}{V_\mu}{B}$ which are conveniently decomposed into the form factors $h_\perp$ and $h_\parallel$ in Heavy Quark Effective Theory (HQET). The computation of the dominant $B^*\pi$ excited states contamination can be carried out in Heavy Meson ChPT (HMChPT) \cite{Wise:1992hn,Burdman:1992gh,Goity:1992tp,Cho:1992cf} and will be presented here. See O. B\"ar's contribution \cite{Bar:lat} for further results on $B^*\pi$ states in HMChPT. We will rely heavily on the material presented there.

\section{Vector Form Factors}

The decay $B \to \pi \ell \bar{\nu}$ can be decomposed into two parts: an initial state $B$ meson decays via the weak interaction into a pion by emitting a $W$ boson. The latter subsequently decays to a $\ell \bar{\nu}$ final state. The propagation and decay of the vector boson can be computed in electroweak perturbation theory, whereas the propagation and decay of the meson via the interaction with the left-handed current $L_\mu = V_\mu - A_\mu$, to which the $W$ boson couples, must be computed in a lattice simulation. Since the axial current does not contribute to the process, the matrix element for $B \to \pi$ can be parametrised as follows:
\begin{equation}
\mel{\pi(p_\pi)}{V_\mu}{B(p_B)} = \left( (p_B + p_\pi)_\mu - q_\mu \frac{m_B^2 - m_\pi^2}{q^2} \right) f_+(q^2) + q_\mu \frac{m_B^2 - m_\pi^2}{q^2} f_0(q^2)\,.
\end{equation}
Here $q_\mu = (p_B - p_\pi)_\mu$ is the momentum transfer. If one assumes $m_\ell = 0$ in the $B \to \pi \ell \bar{\nu}$ decay, only $f_+$ will contribute to the decay width of the process.

In HQET, a possible decomposition reads \cite{Bahr:2019eom}
\begin{align}
&(2 m_{B})^{-1/2} \mel{\pi(p)}{V^0}{B(0)} = h_\parallel(\varv \cdot p)\label{eq:hpar}\,,\\
&(2 m_{B})^{-1/2} \mel{\pi(p)}{V^k}{B(0)} = p^k h_\perp(\varv \cdot p)\label{eq:hperp}\,,
\end{align}
where $\varv^\mu$ is the four velocity of the $B$ meson
\begin{equation}
\varv^\mu = \frac{p^\mu_B}{m_B}\,.
\end{equation}
Matching these form factors to the relativistic ones reveals that $f_+$ is dominated by $h_\perp$ and the contribution of $h_\parallel$ is suppressed by a factor $1/m_B$ \cite{Becirevic:2002sc}.

\section{Parametrization of the Excited States Contamination in the Vector Form Factors}

In lattice simulations, one computes the 3-point function
\begin{equation}
C_{3,\mu} (t,t_\mathrm{v},\vec{p}) = \int\limits_{L^3} \dd[3]{x} \dd[3]{z} e^{- \mathrm{i} \vec{p} (\vec{x} - \vec{z})} \ev{\Pi^+ (t,\vec{x}) V^-_\mu (t_\mathrm{v},\vec{z}) \bar{\mathcal{B}}^{0\dag} (0,\vec{0})}\label{eq:3pt}\,,
\end{equation}
where $\vec{p}$ is the momentum of the final state pion. $\Pi$ and $\bar{\mathcal{B}}$ denote interpolating fields for a pion and a $\bar{B}^0$ meson. Using the 2-point functions $C_2^B$ and $C_2^{\Pi}$ of the interpolating fields, we can define the ratio
\begin{equation}
R_\mu (t,t_\mathrm{v},\vec{p}) = \frac{\sqrt{2 E_\pi} C_{3,\mu} (t,t_\mathrm{v},\vec{p})}{\sqrt{C^B_2 (2 t_\mathrm{v}) C^{\Pi}_2 (2 t - 2 t_\mathrm{v})}}\,,\label{eq:ratio}
\end{equation}
which for $t_\mathrm{v},(t-t_\mathrm{v}) \to \infty$ is equal to the l.h.s. of eqs.~\eqref{eq:hpar} and \eqref{eq:hperp}, respectively, depending on the index $\mu$. Every correlator can be written as the sum of the ground state contribution $C_i^{g.s.}$ and the excited states contamination $C_i^{e.s.}$ which we will use to define
\begin{equation}
C_i = C_i^{g.s.} + C_i^{e.s.} = C_i^{g.s.} \left( 1 + \frac{C_i^{e.s.}}{C_i^{g.s.}} \right) \equiv C_i^{g.s.} (1 + \delta C_i)\,.
\end{equation}
$\delta C_i$ thus quantifies the relative deviation of the correlator from the ground state caused by the excited states contamination. In the following, we will focus on the dominant excited states, which are given by two particle $B^* \pi$ states, see \cite{Bar:lat}. Plugging this decomposition into eq.~\eqref{eq:ratio} and assuming that $\delta C_i$ is small (which is true for large enough time separations), we can Taylor expand to obtain after comparing with eqs.~\eqref{eq:hpar} and \eqref{eq:hperp}
\begin{align}
\delta h_\parallel = \delta C_{3,4} (t,t_\mathrm{v},\vec{p}) - \frac{1}{2} \left( \delta C^B_2 (2 t_\mathrm{v}) + \delta C^{\Pi}_2(2 t - 2 t_\mathrm{v}) \right)\,,\label{eq:exc_hpar}\\
\delta h_\perp = \delta C_{3,k} (t,t_\mathrm{v},\vec{p}) - \frac{1}{2} \left( \delta C^B_2 (2 t_\mathrm{v}) + \delta C^{\Pi}_2(2 t - 2 t_\mathrm{v}) \right)\label{eq:exc_hperp}\,.
\end{align}
The excited states contamination $\delta C^{\Pi}_2$ of the pion 2-point function is much smaller than the other two terms and will be neglected in the following.

Our aim is to compute the above 3-point function in HMChPT to NLO in the chiral expansion and in the static limit. Combining our result for $\delta C_3$ with the result for $\delta C_2^B$ presented in \cite{Bar:lat} allows us to evaluate the excited states contamination of the form factors.

\section{Interpolating Fields}

In order to compute the 3-point function in eq.~\eqref{eq:3pt} in ChPT, we need the interpolating fields and the vector current in the effective theory. These are discussed in \cite{Bar:lat}. The results below will be functions of the NLO Low Energy Constants (LECs) of the (smeared) $B$ meson interpolating field ($\tilde{\beta}_1$) and the vector current ($\beta_1$ and $\beta_2$). Note that if no smearing is applied to the $B$ meson, $\tilde{\beta}_1 = \beta_1$ due to Heavy Quark Spin Symmetry. A strategy to determine $\beta_1$ can be found in \cite{Bar:lat}, the determination of $\beta_2$ is presented in section \ref{sec:b2}.

\begin{figure}[]
\centering
\subfloat[]{
	\begin{tikzpicture}

	\draw (0,0) -- (0.2,0) -- (0.1,{sqrt(0.04)}) -- (0,0);
	\draw[fill] (2,0) -- (2.2,0) -- (2.1,{sqrt(0.04}) -- (2,0);

	\draw[double,->-] (0.15,0.1) -- (2.05,0.1);
	
	\draw[dashed] (2.1,0.1) arc(180:0:0.9 and 0.7);
	\draw(3.9,0.1) node[cross=3pt] {};
	
	\end{tikzpicture}
}
\hfil
\subfloat[]{
	\begin{tikzpicture}
	
	\draw (0,0) -- (0.2,0) -- (0.1,{sqrt(0.04)}) -- (0,0);
	\draw[fill] (2,0) -- (2.2,0) -- (2.1,{sqrt(0.04}) -- (2,0);

	\filldraw (1,0.1) circle (0.07);
	\draw (3.9,0.1) node[cross=3pt] {};

	\draw[double,->-] (0.15,0.1) -- (0.92,0.1);
	\draw[double,->-] (1.08,0.1) -- (2.05,0.1);
	\draw [dashed](1,0.1) arc(180:0:1.45 and 0.7);
	
	\end{tikzpicture}
}
\hfil
\subfloat[]{
	\begin{tikzpicture}
	
	\draw (0,0) -- (0.2,0) -- (0.1,{sqrt(0.04)}) -- (0,0);
	\draw[fill] (2,0) -- (2.2,0) -- (2.1,{sqrt(0.04}) -- (2,0);

	\draw (3.9,0.2) node[cross=3pt] {};
	\draw [dashed](0.1,0.2) arc(180:0:1.9 and 0.7);

	\draw[double,->-] (0.15,0.1) -- (2.05,0.1);
	
	\end{tikzpicture}
}
\hfil
\subfloat[]{
	\begin{tikzpicture}

	\draw (0,0) -- (0.2,0) -- (0.1,{sqrt(0.04)}) -- (0,0);
	\draw[fill] (2,0) -- (2.2,0) -- (2.1,{sqrt(0.04}) -- (2,0);

	\draw[double,->-] (0.15,0.1) -- (2.05,0.1);
	
	\draw [dashed](2.1,0.1) arc(180:0:0.9 and 0.7);
	\draw(3.9,0.1) node[cross=3pt] {};
	\draw[dashed] (0.1,0.2) arc(180:0:1 and 0.5);
	
	\end{tikzpicture}
}
\hfil
\subfloat[]{
\begin{tikzpicture}

	\draw (0,0) -- (0.2,0) -- (0.1,{sqrt(0.04)}) -- (0,0);
	\draw[fill] (2,0) -- (2.2,0) -- (2.1,{sqrt(0.04}) -- (2,0);

	\draw[double,->-] (0.15,0.1) -- (0.92,0.1);
	\draw[double,->-] (1.08,0.1) -- (2.05,0.1);
	\filldraw (1,0.1) circle (0.07);
	
	\draw [dashed](2.1,0.1) arc(180:0:0.9 and 0.5);
	\draw[dashed] (1,0.2) arc(180:0:0.55 and 0.5);
	\draw(3.9,0.1) node[cross=3pt] {};
	
	\end{tikzpicture}
}
\hfil
\subfloat[]{
\begin{tikzpicture}

	\draw (0,0) -- (0.2,0) -- (0.1,{sqrt(0.04)}) -- (0,0);
	\draw[fill] (3,0) -- (3.2,0) -- (3.1,{sqrt(0.04}) -- (3,0);

	\draw[double,->-] (0.15,0.1) -- (0.92,0.1);
	\draw[double,->-] (1.08,0.1) -- (1.92,0.1);
	\draw[double,->-] (2.08,0.1) -- (3.05,0.1);
	\filldraw (1,0.1) circle (0.07);
	\filldraw (2,0.1) circle (0.07);
	
	\draw [dashed](3.1,0.1) arc(180:0:0.4 and 0.45);
	\draw[dashed] (1,0.2) arc(180:0:0.5 and 0.45);
	\draw(3.9,0.1) node[cross=3pt] {};
	
	\end{tikzpicture}
}
\hfil
\subfloat[]{
\begin{tikzpicture}

	\draw (0,0) -- (0.2,0) -- (0.1,{sqrt(0.04)}) -- (0,0);
	\draw[fill] (2,0) -- (2.2,0) -- (2.1,{sqrt(0.04}) -- (2,0);

	\draw[double,->-] (0.15,0.1) -- (0.92,0.1);
	\draw[double,->-] (1.08,0.1) -- (2.05,0.1);
	\filldraw (1,0.1) circle (0.07);
	
	\draw [dashed](2.1,0.1) arc(180:0:0.9 and 0.7);
	\draw[dashed] (0.1,0.2) arc(180:0:0.45 and 0.45);
	\draw(3.9,0.1) node[cross=3pt] {};
	
	\end{tikzpicture}
}
\hfil
\subfloat[]{
\begin{tikzpicture}

	\draw (0,0) -- (0.2,0) -- (0.1,{sqrt(0.04)}) -- (0,0);
	\draw[fill] (2,0) -- (2.2,0) -- (2.1,{sqrt(0.04}) -- (2,0);

	\draw[double,->-] (0.15,0.1) -- (0.92,0.1);
	\draw[double,->-] (1.08,0.1) -- (2.05,0.1);
	\filldraw (1,0.1) circle (0.07);
	
	\draw [dashed](1,0.1) arc(180:0:1.45 and 0.7);
	\draw[dashed] (0.1,0.2) arc(180:0:0.45 and 0.45);
	\draw(3.9,0.1) node[cross=3pt] {};
	
	\end{tikzpicture}
}
\hfil
\subfloat[]{
\begin{tikzpicture}

	\draw (0,0) -- (0.2,0) -- (0.1,{sqrt(0.04)}) -- (0,0);
	\draw[fill] (3,0) -- (3.2,0) -- (3.1,{sqrt(0.04}) -- (3,0);

	\draw[double,->-] (0.15,0.1) -- (0.92,0.1);
	\draw[double,->-] (1.08,0.1) -- (1.92,0.1);
	\draw[double,->-] (2.08,0.1) -- (3.05,0.1);
	\filldraw (1,0.1) circle (0.07);
	\filldraw (2,0.1) circle (0.07);
	
	\draw [dashed](2,0.1) arc(180:0:0.95 and 0.45);
	\draw[dashed] (1,0.2) arc(180:0:0.5 and 0.45);
	\draw(3.9,0.1) node[cross=3pt] {};
	
	\end{tikzpicture}
}
\hfil
\subfloat[]{
\begin{tikzpicture}

	\draw (-1,0) -- (-0.8,0) -- (-0.9,{sqrt(0.04)}) -- (-1,0);
	\draw[fill] (3,0) -- (3.2,0) -- (3.1,{sqrt(0.04}) -- (3,0);

	\draw[double,->-] (-0.85,0.1) -- (0.15,0.1);
	\draw[double,->-] (0.15,0.1) -- (0.92,0.1);
	\draw[double,->-] (1.08,0.1) -- (1.92,0.1);
	\draw[double,->-] (2.08,0.1) -- (3.05,0.1);
	\filldraw (1,0.1) circle (0.07);
	\filldraw (2,0.1) circle (0.07);
	\filldraw (0,0.1) circle (0.07);
	
	\draw [dashed](2,0.1) arc(180:0:0.95 and 0.45);
	\draw[dashed] (0,0.2) arc(180:0:0.5 and 0.45);
	\draw(3.9,0.1) node[cross=3pt] {};
	
	\end{tikzpicture}
}
\hfil
\subfloat[]{
\begin{tikzpicture}

	\draw (0,0) -- (0.2,0) -- (0.1,{sqrt(0.04)}) -- (0,0);
	\draw[fill] (3,0) -- (3.2,0) -- (3.1,{sqrt(0.04}) -- (3,0);

	\draw[double,->-] (0.15,0.1) -- (0.92,0.1);
	\draw[double,->-] (1.08,0.1) -- (1.92,0.1);
	\draw[double,->-] (2.08,0.1) -- (3.05,0.1);
	\filldraw (1,0.1) circle (0.07);
	\filldraw (2,0.1) circle (0.07);
	
	\draw [dashed](2,0.1) arc(180:0:0.95 and 0.45);
	\draw[dashed] (0.1,0.2) arc(180:0:0.45 and 0.45);
	\draw(3.9,0.1) node[cross=3pt] {};
	
	\end{tikzpicture}
}
\hfil
\subfloat[]{
\begin{tikzpicture}

	\draw (0,0) -- (0.2,0) -- (0.1,{sqrt(0.04)}) -- (0,0);
	\draw[fill] (3,0) -- (3.2,0) -- (3.1,{sqrt(0.04}) -- (3,0);

	\draw[double,->-] (0.15,0.1) -- (0.92,0.1);
	\draw[double,->-] (1.08,0.1) -- (1.92,0.1);
	\draw[double,->-] (2.08,0.1) -- (3.05,0.1);
	\filldraw (1,0.1) circle (0.07);
	\filldraw (2,0.1) circle (0.07);
	
	\draw [dashed](2,0.1) arc(180:0:0.95 and 0.45);
	\draw[dashed] (1,0) arc(-180:0:1.05 and 0.45);
	\draw(3.9,0.1) node[cross=3pt] {};
	
	\end{tikzpicture}
}
\hfil
\subfloat[]{
\begin{tikzpicture}

	\draw (0,0) -- (0.2,0) -- (0.1,{sqrt(0.04)}) -- (0,0);
	\draw[fill] (3,0) -- (3.2,0) -- (3.1,{sqrt(0.04}) -- (3,0);

	\draw[double,->-] (0.15,0.1) -- (0.92,0.1);
	\draw[double,->-] (1.08,0.1) -- (1.92,0.1);
	\draw[double,->-] (2.08,0.1) -- (3.05,0.1);
	\filldraw (1,0.1) circle (0.07);
	\filldraw (2,0.1) circle (0.07);
	
	\draw [dashed](2,0.1) arc(180:0:0.95 and 0.45);
	\draw[dashed] (0.1,0) arc(-180:0:1.5 and 0.45);
	\draw(3.9,0.1) node[cross=3pt] {};
	
	\end{tikzpicture}
}

\caption{Feynman diagrams for the 3-point function $C_{3,\mu}$. The cross represents the interpolating field destroying the pion, the filled triangle the vector current, and the triangle on the left denotes the interpolating field creating a $\bar{B}^0$. Double lines represent heavy meson propagators, the dashed lines the propagator of the pion. We do not distinguish between the pseudoscalar $B$ and vector $B^*_\mu$ meson propagator.}
\label{fig:diagrams_ff}
\end{figure}
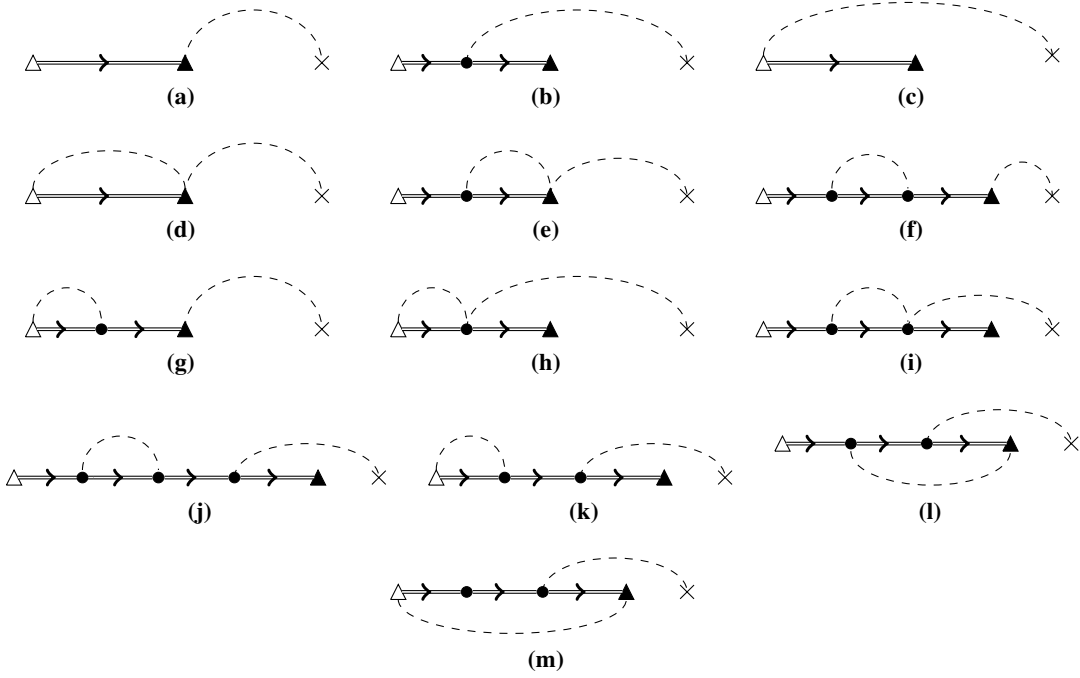

\section{Results}

\subsection{$h_\perp$}

It follows from eqs.~\eqref{eq:exc_hpar} and \eqref{eq:exc_hperp} that we need the excited states of the $B$ meson 2-point function for both form factors. The result can be found in \cite{Bar:lat}. For the 3-point function with spatial components of the vector current, the result can be parametrised as follows  (see Fig.~\ref{fig:diagrams_ff} for the Feynman diagrams):
\begin{equation}
\delta C_{3,k} (t,t_\mathrm{v},\vec{p}) = - \frac{1 + \tilde{\beta}_1 E_\pi(\vec{p})/\varg}{1 - \beta_1 E_\pi(\vec{p})/\varg} e^{- E_\pi(\vec{p}) t_\mathrm{v}} + \sum \limits_{\vec{l}} \frac{1}{(f L)^2 (E_\pi(\vec{l}) L)} c(\vec{l},\vec{p},\beta_1,\tilde{\beta}_1,\gamma) e^{- E_\pi(\vec{l}) t_\mathrm{v}}\,.\label{eq:d_c3_k}
\end{equation}
The first term stands out since it is not volume-suppressed, i.e. there is no factor $1/L^3$ and no momentum sum. This correlator is the only one considered here and in \cite{Bar:lat} with such a term, but it has already been observed in induced nucleon form factors in \cite{Bar:2018xyi} and there explained the large excited states contamination in the lattice correlator. Diagrams (b) and (c) in Fig.~\ref{fig:diagrams_ff} are responsible for this ``volume-enhanced'' contribution. Setting $\beta_1 = \tilde{\beta}_1 = 0$, i.e. at leading order in the chiral expansion, we see that this term is simply given by the negative of an exponential, which falls off with the energy of the outgoing pion. One can thus expect to find a large negative excited states contamination since the second term in eq.~\eqref{eq:d_c3_k} and the excited states contribution of the 2-point function are expected to be smaller.

\begin{figure}[!t]
\centering
\includegraphics{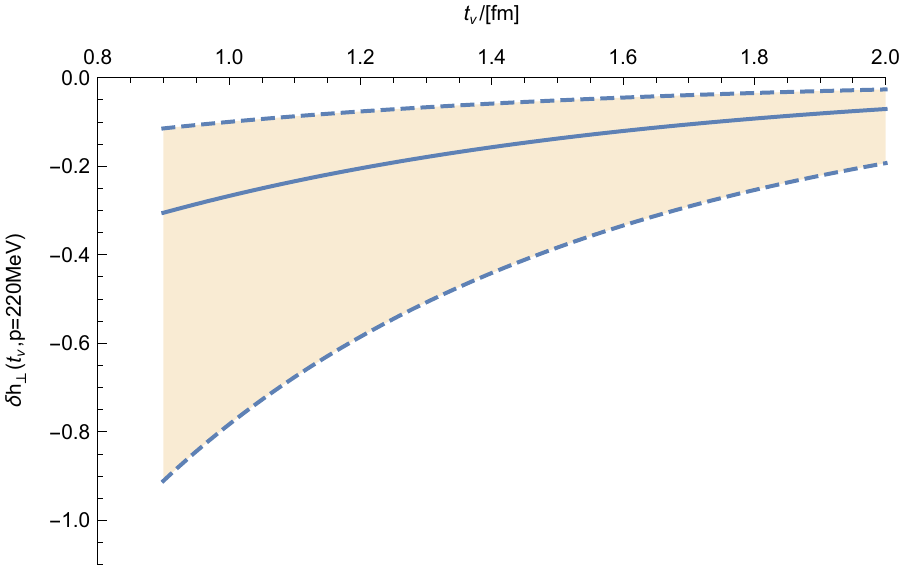}
\caption{The LO deviation $\delta h_\perp$ (solid line) and bounds for the NLO estimates (dashed lines) for a pion momentum of $\abs{\vec{p}} = 220$~MeV as a function of $t_\mathrm{v}$. \label{fig:d_f_k}}
\end{figure}
\begin{figure}[!t]
\centering
\includegraphics{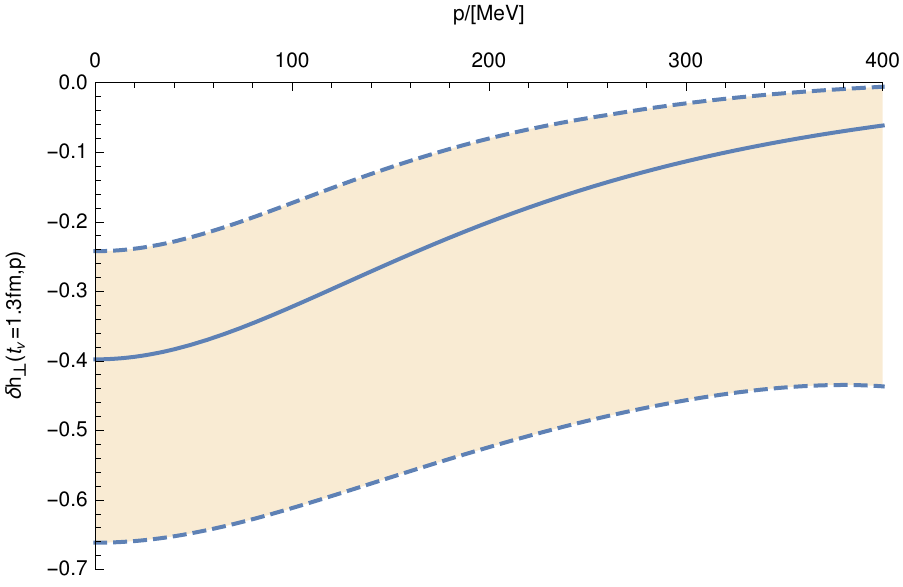}
\caption{The LO deviation $\delta h_\perp$ (solid line) and bounds for the NLO estimates (dashed lines) at $t_\mathrm{v} =1.3$~fm as a function of $\abs{\vec{p}}$. \label{fig:d_f_k_p}}
\end{figure}

The coefficient function $c$ of the volume suppressed contribution is a function of the LECs of the interpolating field and vector current, $\tilde{\beta}_1$ and $\beta_1$, and $\gamma$, which is a linear combination of LECs of the NLO Lagrangian. It is the same combination which was already found for the excited states contamination of the $B B^* \pi$ coupling \cite{Bar:lat}.

Fig.~\ref{fig:d_f_k} shows the result of $\delta h_\perp$ as a function of $t_\mathrm{v}$, the timespan for which the $B^*\pi$ excited state propagates, and a fixed final state pion momentum of $\vec{p} = 220$~MeV, whereas Fig.~\ref{fig:d_f_k_p} fixes $t_\mathrm{v} = 1.3$~fm and varies $\vec{p}$. The momentum of 220~MeV corresponds to the smallest non-zero momentum on a lattice with size $m_\pi L = 4$ with our parameters. The plots were created by taking the infinite volume limit of the formulae, fixing $m_\pi = 140$~MeV, $f = 93$~MeV, and $\varg = 0.5$. Since numerical estimates for the NLO LECs $\beta_1$, $\tilde{\beta}_1$, and $\gamma$ do not exist, we vary them in the range $[- \Lambda_\chi^{-1}, \Lambda_\chi^{-1}]$ which can be obtained from dimensional analysis. Here $\Lambda_\chi = 4 \pi f$ denotes the chiral symmetry breaking scale. It is also important to note that ChPT is a low energy effective theory and does not account for the high energy pions in the momentum sums/integrals. In order to suppress them, one should consider the results only for $t_\mathrm{v} \gtrsim 1.3$~fm.

As is evident from the figures, the excited states contribution are negative and large. This is mostly due to the first term in eq.~\eqref{eq:d_c3_k}, the second term is of the order of a few percent. This also explains why the excited states decrease as a function of $\vec{p}$, since the exponent of the first term is proportional to $E_\pi(\vec{p})$. Overall, one expects large excited states for $h_\perp$ which might lower the lattice results significantly if not taken care of accordingly.

\subsection{$h_\parallel$}

\begin{figure}[!t]
\centering
\includegraphics{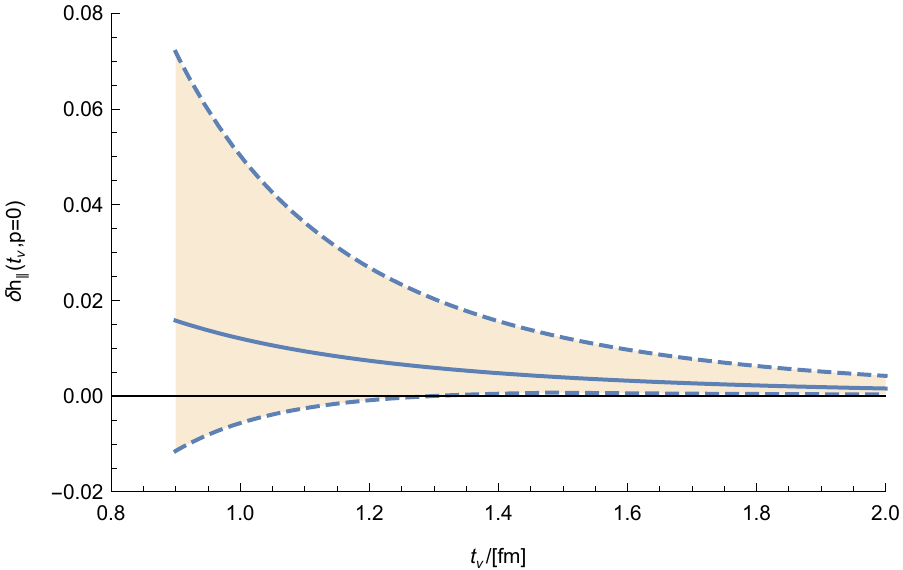}
\caption{The LO deviation $\delta h_\parallel$ (solid line) and bounds for the NLO estimates (dashed lines) for vanishing final state pion momentum as a function of $t_\mathrm{v}$. \label{fig:d_f_4}}
\end{figure}
\begin{figure}[!t]
\centering
\includegraphics{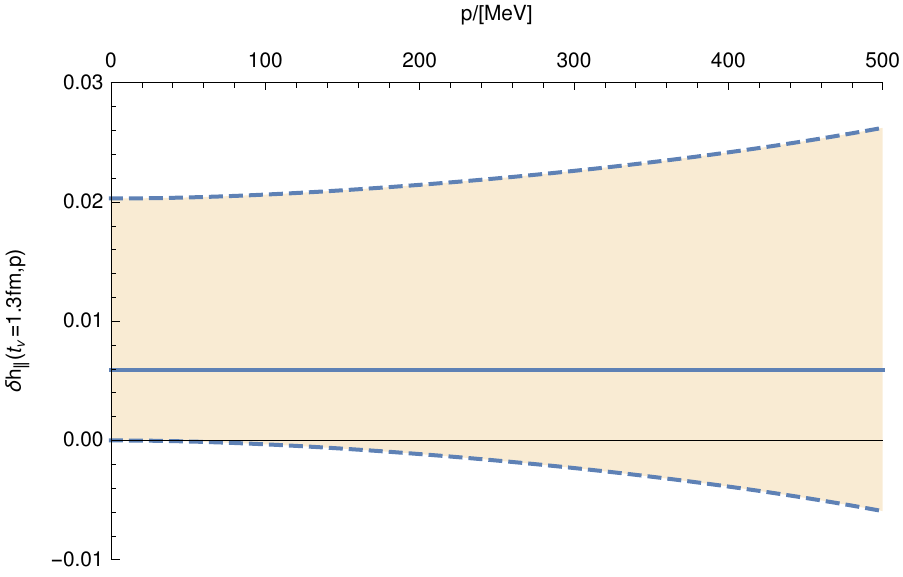}
\caption{The LO deviation $\delta h_\parallel$ (solid line) and bounds for the NLO estimates (dashed lines) at $t_\mathrm{v} =1.3$~fm as a function of $\abs{\vec{p}}$. \label{fig:d_f_4_p}}
\end{figure}

The 3-point function with the temporal component of the vector current, $V_4$, can be parametrised as
\begin{equation}
\delta C_{3,4} (t,t_\mathrm{v},\vec{p}) = \sum \limits_{\vec{l}} \frac{1}{(f L)^2 (E_\pi(\vec{l}) L)} d(\vec{l},\vec{p},\beta_1,\tilde{\beta}_1,\beta_2) e^{- E_\pi(\vec{l}) t_\mathrm{v}}\,.\label{eq:d_c3_4}
\end{equation}
In contrast to $\delta h_\perp$, the function $d$ is a function of both NLO LECs of the vector current: $\beta_1$ and $\beta_2$. A strategy to compute the latter in a lattice simulation will be presented in the next section. We see that there is only a volume-suppressed contribution which suggests a smaller excited state contamination for $h_\parallel$ than for $h_\perp$.

Figs.~\ref{fig:d_f_4} and \ref{fig:d_f_4_p} show the results for $\delta h_\parallel$ as a function of $t_\mathrm{v}$ and $\vec{p}$, respectively. In contrast to $\delta h_\perp$, the excited states can be positive or negative and are only of the order of a few percent. Furthermore, the dependence on the final state pion momentum $\vec{p}$ is mild. The impact on lattice simulations is thus expected to be small.

\section{Determination of $\beta_2$}\label{sec:b2}

The results for $\delta h_\parallel$ and $\delta h_\perp$ depend on the LECs $\beta_1$ and $\beta_2$ of the interpolating fields and vector current. We now present a method to determine $\beta_2$ which is similar to the one for $\beta_1$ presented in \cite{Bar:lat}. It is necessary to compute the 3-point function
\begin{equation}
C_3(t,t_A,\vec{q}) = \int\limits_{L^3} \dd[3]{x} \dd[3]{y} e^{\mathrm{i} \vec{q} \vec{x}} \ev{A_4^{ll}(t,\vec{x}) A_k^{hl}(t_A,\vec{z}) \bar{\mathcal{B}}^{*\dag}_k (0,\vec{0})}\,,
\end{equation}
where $A^{ll}_4$ is the time component of the light-light axial vector current (destroying a pion), and $A^{hl}_k$ is a spatial component of the heavy-light axial current. Note the injection of momentum for the pion. Computing the correlator to LO in ChPT (the diagram looks like (a) in Fig.~\ref{fig:diagrams_ff}) and dividing by the $B$ meson 2-point function yields
\begin{equation}
R(t,t_A,\vec{q}) = - 2 \mathrm{i} \frac{C_3(t,t_A,\vec{q})}{C_2^B(t_A)} = (1 - \beta_2 E_\pi(\vec{q})) e^{- (t - t_A) E_\pi(\vec{q})}\,.
\end{equation}
The excited states contamination in the 3-point function is suppressed by a relative factor $\exp(- t_A E_\pi)$. Here, we have assumed that the 2-point function was computed with one point-like and one smeared interpolating field. $\beta_2$ can then be determined by computing $R$ for several values of the pion momentum $\vec{q}$. The time dependence can be eliminated by determining $E_\pi(\vec{q})$ and multiplying the ratio with $\exp((t - t_A) E_\pi (\vec{q}))$. $\beta_2$ can then be determined by a linear fit in $E_\pi(\vec{q})$.

\section{Conclusions}

We have computed the excited states contamination in the form factors $h_\parallel$ and $h_\perp$ in the static limit of HMChPT. The relative magnitude of the excited states is of the order of a few percent for $h_\parallel$ and has only a mild dependence on the momentum of the final state pion, $\vec{p}$. On the other hand, $h_\perp$ receives large negative contributions which are more severe for smaller $\vec{p}$. This is due to the ``tree level'' diagrams (b) and (c) in Fig.~\ref{fig:diagrams_ff} which are analogous to diagrams computed for the induced nucleon form factors in \cite{Bar:2018xyi}.

Considering the results presented here and in \cite{Bar:lat}, there are in total three unknown LECs: $\beta_1$, $\beta_2$, and $\gamma$ (since $\tilde{\beta}_1$ can be determined in the same way as $\beta_1$, we consider them as one LEC here). A promising procedure for determining $\gamma$ is not known to us, but our results do not depend on it significantly. The most important LEC is the ubiquitous $\beta_1$ which shows up in every single correlator. A determination of this LEC in a lattice simulation is thus essential in order to have reliable estimate for the excited states contamination of $B$ meson observables. We can nevertheless state that (with the exception of $\delta h_\perp$) all observables considered so far are at most of the order 10\% at $t \approx 1.3$~fm and lead to an overestimation.

\vspace{4ex}
\noindent {\bf Acknowledgments}
AB’s research is funded by the Deutsche Forschungsgemeinschaft (DFG, German Research Foundation), Projektnummer 417533893/GRK2575 “Rethinking Quantum Field Theory”.
\vspace{3ex}

\bibliographystyle{JHEP}
\bibliography{lattice2022_abroll}

\end{document}